\documentclass[floats,floatfix,showpacs,amssymb,prd,twocolumn,superscriptaddress,nofootinbib,nolongbibliography,reprint]{revtex4-2}

\usepackage{amssymb,amsmath,verbatim,mathtools,needspace,enumitem,etoolbox,graphicx,physics,microtype,afterpage}
\usepackage{gensymb}

\usepackage[dvipsnames, usenames]{xcolor}

\definecolor{linkcolor}{rgb}{0.0,0.3,0.5}
\usepackage[unicode, colorlinks=true, linkcolor=linkcolor, citecolor=linkcolor, filecolor=linkcolor,urlcolor=linkcolor, pdfusetitle]{hyperref}
\usepackage{textcomp}
\usepackage[all]{hypcap}
\usepackage[T1]{fontenc}
\usepackage[utf8]{inputenc}
\usepackage{tabularx}
\usepackage[normalem]{ulem}
\usepackage{placeins}
\usepackage{float}
\usepackage{verbatim}
\usepackage{graphicx}
\usepackage {tikz}
\usetikzlibrary {positioning}
\definecolor {processblue}{cmyk}{0.96,0,0,0}

\newcolumntype{P}[1]{>{\centering\arraybackslash}p{#1}}
\renewcommand{\vec}[1]{\mathbf{#1}}

\def\apjl{\rm{Astrophys.~J.~ Lett.}}
\def\apjs{\rm{ApJS}}
\def\aap{\rm{Astron.~Astrophys.}}

\def\mnras{\rm{Mon.~Not.~R.~Astron.~Soc.}}
\def\prd{\rm{Phys.~Rev.~D}}

\newcommand{\dallas}{\affiliation{Department of Physics, The University of Texas at Dallas, Richardson, Texas 75080, USA}}

\usepackage [english]{babel}
\usepackage [autostyle, english = american]{csquotes}
\MakeOuterQuote{"}

\newcommand\orcid[1]{\href{https://orcid.org/#1}{$\!$\includegraphics[scale=0.006]{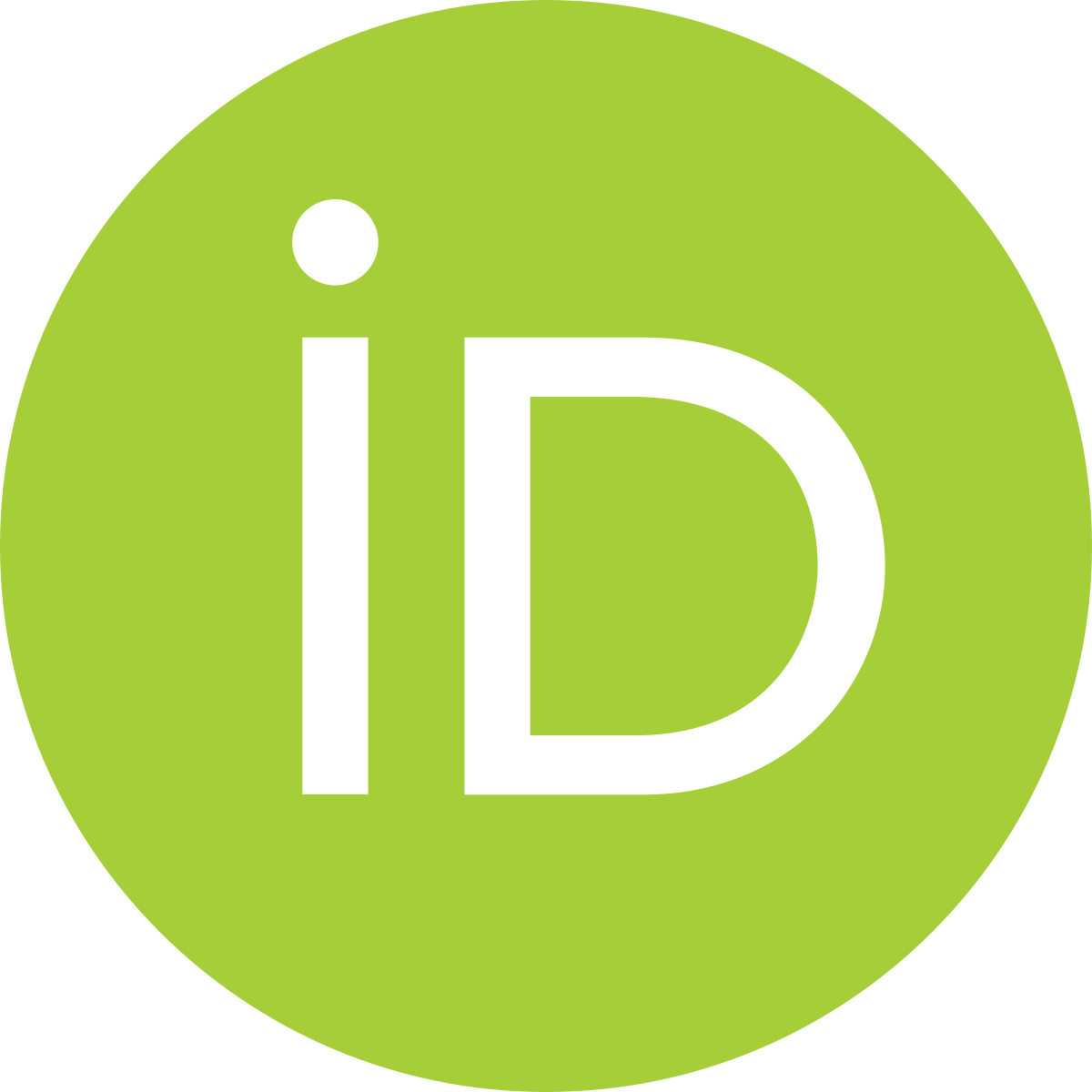} $\!\!$}}

\begin{document}

\title{Detectability of strongly lensed gravitational waves using model-independent image parameters}

\author{Saif Ali \orcid{0000-0002-6971-4971}}
\email{sxa180025@utdallas.edu}
\dallas

\author{Evangelos Stoikos \orcid{0000-0002-1043-3673}}
\email{Evangelos.Stoikos@utdallas.edu}
\dallas

\author{Evan Meade \orcid{0000-0001-9854-4812}}
\dallas

\author{Michael Kesden \orcid{0000-0002-5987-1471}}
\email{kesden@utdallas.edu}
\dallas

\author{Lindsay King \orcid{0000-0001-5732-3538}}
\email{lindsay.king@utdallas.edu}
\dallas

\date{\today}

\begin{abstract}

Strong gravitational lensing of gravitational waves (GWs) occurs when the GWs from a compact binary system travel near a massive object. The mismatch between a lensed signal and unlensed templates determines whether lensing can be identified in a particular GW event. For axisymmetric lens models, the lensed signal is traditionally calculated in terms of model-dependent lens parameters such as the lens mass $M_L$ and source position $y$. We propose that it is useful to parameterize this signal instead in terms of model-independent image parameters: the flux ratio $I$ and time delay $\Delta t_d$ between images. The functional dependence of the lensed signal on these image parameters is far simpler, facilitating data analysis for events with modest signal-to-noise ratios.  In the geometrical-optics approximation, constraints on $I$ and $\Delta t_d$ can be inverted to constrain $M_L$ and $y$ for any lens model including the point mass (PM) and singular isothermal sphere (SIS) that we consider.  We use our model-independent image parameters to determine the detectability of gravitational lensing in GW signals and find that for GW events with signal-to-noise ratios $\rho$ and total mass $M$, lensing should in principle be identifiable for flux ratios $I \gtrsim 2\rho^{-2}$ and time delays $\Delta t_d \gtrsim M^{-1}$.

\end{abstract} 

\maketitle
\graphicspath{ {./Figures} }

\section{\label{sec:Introduction} Introduction}

The first direct detection of gravitational waves (GWs) from merging compact objects was reported by the LIGO and Virgo collaborations in 2016 \cite{2016}. To date, Advanced LIGO \cite{LIGOScientific:2014pky} and Advanced Virgo \cite{VIRGO:2014yos} have reported about 90 events, most of which are mergers between stellar-mass black holes, during their first three observing runs \cite{abbott2021gwtc-3}. Kamioka Gravitational Wave Detector (KAGRA) \cite{Somiya:2011np, PhysRevD.88.043007, KAGRA:2020tym} has joined the preexisting ground-based GW detectors to form the Advanced LIGO-Virgo-Kagra (LVK) network. The increased sensitivity of detectors such as LVK has allowed us to detect an increasing number of GW events and to perform various general relativistic and cosmological tests \cite{abbott2021tests, ligo2021constraints}. With the increasing sensitivity of the current ground-based detector network and future detectors such as the Cosmic Explorer (CE) \cite{evans2021horizon}, the Einstein Telescope (ET) \cite{Maggiore:2019uih}, the Deci-Hertz Interferometer Gravitational Wave Observatory (DECIGO) \cite{Kawamura:2020pcg}, and the Laser Interferometer Space Antenna (LISA) \cite{Barausse:2020rsu}, the number of observed GW events will increase dramatically, as will the probability of observing new propagation effects such as gravitational lensing that have yet to be detected \cite{abbott2021search}. 

When GWs travelling through the Universe encounter a massive object, such as a compact object, galaxy or galaxy cluster, that can act as a lens, deflection of these GWs, i.e. gravitational lensing, will occur \cite{1971NCimB...6..225L, 1974IJTP....9..425O, 1999PThPS.133..137N, PhysRevLett.80.1138, Takahashi_2003,Oguri:2018muv, Li:2018prc}. Strong lensing of GWs will arise when a lens is very close to the line of sight. This will result in the GWs splitting into different lensed images, each with its own magnification and phase \cite{Takahashi_2003, Ezquiaga_2021}. There will also be an associated time delay between the lensed images which could range from seconds to years depending on the mass of the lens and geometry of the lens system \cite{1985A&A...143..413S, 1992grle.book.....S}.
 
GW lensing, if detected, could facilitate several exciting scientific studies. It could be used to extract information about the existence of intermediate-mass (mass ranging from $\sim 10^2$ - $10^5 M_\odot$) \cite{Lai:2018rto} or primordial black holes \cite{Diego:2019rzc, Oguri:2020ldf} and test general relativity \cite{Baker:2016reh, Collett:2016dey, Mukherjee:2019wcg}, including through constraints from GW polarization content \cite{PhysRevD.103.024038}. In addition, if a lensed electromagnetic (EM) counterpart of the lensed GW event is observed, it could help to locate the host galaxy at sub-arcsecond precision \cite{Hannuksela:2020xor}. Combining the information from the two messengers, i.e. GW and EM lensing, could enable high-precision cosmography \cite{Sereno:2011ty, Liao:2017ioi, Cao:2019kgn, Li:2019rns, Yu:2020agu, wempe2022lensing}.

There are two major differences between the gravitational lensing of EM waves and GWs from the point of view of wave-optics effects. The first difference is in the applicability of the geometrical-optics approximation.
In the case of EM waves, this approximation, typically valid when the wavelength $\lambda$ of the waves is much smaller than the Schwarzchild radius $R_s$ of the lens, applies to the vast majority of observations. This is not always the case for GWs, since ground-based detectors such as the LVK network observe at frequencies $(10$ --- $10^4)$ Hz, lower than even the lowest-frequency radio telescopes. These GWs have wavelengths longer than the Schwarzschild radii of lenses with masses $M_L \lesssim 10^4 M_\odot$, leading to non-negligible wave-optics effects. The second difference is that the GWs emitted by compact binaries, unlike most EM sources, are coherent, causing interference between lensed images when the signals overlap at the observer.

In this paper, we focus on strong gravitational lensing by stellar-mass objects and GW sources consistent with those seen by the LVK network. However, our treatment is also applicable to more massive lenses, and to sources such as supermassive binary black holes that will be detectable by LISA. In the frequency domain, the modulation of GWs due to gravitational lensing is characterized by a multiplicative factor known as the amplification factor. Typically, this factor is parameterized in terms of \emph{model-dependent} lens parameters such as the source position $y$ and lens mass $M_L$ \cite{1999PThPS.133..137N, Takahashi_2003}. In the limit where the geometrical-optics approximation is valid, for a particular axisymmetric lens model, analytical equations relate these lens parameters to a set of \emph{model-independent} image parameters, the flux ratio $I$ and time delay $\Delta t_d$ between the images. Motivated by current GW search pipelines which use unlensed GW templates, we explore the detectability of lensing signatures using a match-filtering analysis between lensed GW source and unlensed GW templates. For the axisymmetric lens models, we explore the mismatch between the lensed and unlensed GW waveforms in both the lens and image parameter spaces.

This paper is organized as follows. In Sec.~\ref{sec:basic formalism}, we begin with a pedagogical outline of gravitational lensing of GWs, discussing the time delay and amplification factor due to the lens. We then present the prescription used to generate the GWs in the inspiral phase of binary compact objects using the post-Newtonian approximation following \cite{PhysRevD.49.2658}. In Sec.~\ref{sec:axisymmetric lens}, we present a detailed analysis of the  point mass (PM) and singular isothermal sphere (SIS) axisymmetric lens-mass profiles and introduce the model-independent image parameters. In Sec.~\ref{sec: Match filtering analysis}, we perform a match-filtering analysis in which we calculate the mismatch between lensed and unlensed GWs. Appendices~\ref{appendix:Mismatch approximation in geometrical-optics limit} and \ref{appendix:Investigating striations in the mismatch} investigate the mismatch between lensed GW source and unlensed templates. Throughout the paper, we assume $c = G = 1$.

\section{\label{sec:basic formalism} Basic formalism }

In this section, we briefly review the basic theory of the gravitational lensing of GWs.

\subsection{\label{subsec:gravitational lensing}Gravitational lensing}

\begin{figure}[t!]
    \centering
    \includegraphics[scale = 0.52]{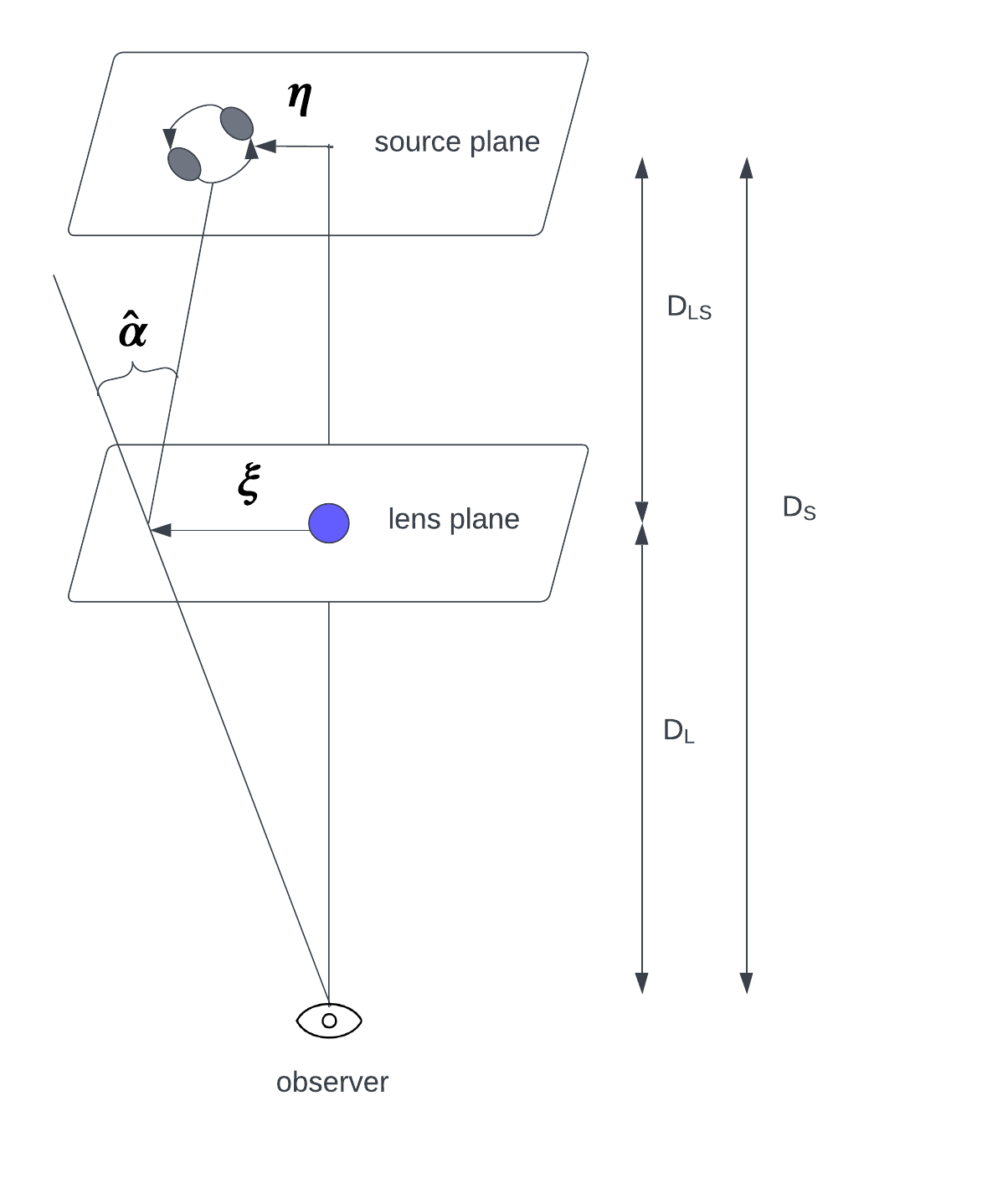}
    \caption{A typical gravitational lens system consisting of a compact binary system in the source plane, a lens in the lens plane, and an observer.  $D_L$, $D_{LS}$ and $D_S$ are the angular-diameter distances from observer to lens, lens to source, and observer to source respectively. The vector $\boldsymbol{\xi}$ is the impact parameter in the lens plane, and the vector $\boldsymbol{\eta}$ is the location of the source with respect to the optic axis in the source plane.  $\hat{\boldsymbol{\alpha}}$ is the deflection angle measured on the lens plane.}
    \label{fig:lens system}
\end{figure}

In the strong gravitational-lensing regime, we observe multiple images (or a single very distorted image) of a distant background source due to the presence of an intervening massive astrophysical object known as a lens. Lensing occurs when the GWs from a compact binary system travel near a lens as shown for the general lensing geometry in Fig.~\ref{fig:lens system} \cite{1992grle.book.....S}. The extents of the lens and the source are taken to be much less than the observer-lens and lens-source distances, in which case they can be localized to the lens and source planes. An optic axis connects the observer and the center of the lens. The lens and source planes are at angular-diameter distances $D_L$ and $D_S$ respectively. The angular-diameter distance between the lens and source planes is $D_{LS}$. A GW source is located on the source plane at displacement $\boldsymbol{\eta}$ with respect to the optic axis. After being emitted by the source, the GWs travel to the lens plane, with an impact parameter $\boldsymbol{\xi}$, and are deflected through an angle $\hat{\boldsymbol{\alpha}}$ by the gravitational potential of the lens. $\vec{x} \equiv \boldsymbol{\xi} / \xi_0$ and $\vec{y} \equiv \boldsymbol{\eta}/ (\xi_0 D_S / D_L)$ are dimensionless vectors on the lens and source plane respectively, where $\xi_0$ is a model-dependent characteristic length scale on the lens plane called the Einstein radius. GWs that reach the observer satisfy the lens equation
\begin{equation} \label{eq:lens equation}
\vec{y} = \vec{x} -\boldsymbol{\alpha}(\vec{x})\,,
\end{equation}
where
\begin{equation}
\boldsymbol{\alpha}(\vec{x}) = \frac{D_L D_{LS}}{\xi_0 D_S} \boldsymbol{\hat{\alpha}}(\xi_0 \vec{x}) = \nabla_x \psi(\vec{x}) \,,
\end{equation}
is the scaled deflection angle at the observer. The lensing potential $\psi(\vec{x})$ is given by the two-dimensional Poisson equation
\begin{equation} \label{E:Poisson}
\nabla_x^2 \psi(\vec{x}) = \frac{2\Sigma(\vec{x})}{\Sigma_{\rm cr}}~,
\end{equation}
where $\Sigma$ is the surface mass density of the lens and $\Sigma_{\rm cr} \equiv D_S/4\pi D_LD_{LS}$ is the critical surface mass density. For the formation of multiple images, $\Sigma/\Sigma_{\rm cr} > 1$ is a sufficient, but not necessary condition \cite{1986MNRAS.219..333S}. 

Gravitational lensing causes a time delay between the lensed images at the observer. The arrival time has two components, one arising from the geometry of the path traveled, and the other due to the gravitational potential of the lens known as the Shapiro time delay. The time delay at the observer due to a lens at redshift $z_L$ is
\begin{align}\label{eq:time delay}
    t_d(\vec{x}, \vec{y}) = \frac{D_S \xi^2_0 (1 + z_L) }{D_L D_{LS}}\left[ \frac{1}{2} |\vec{x} - \vec{y}|^2 - \psi(\vec{x}) + \phi_m(\vec{y}) \right],
\end{align}
where $\phi_m(\vec{y})$ is chosen such that the minimum value of the time delay is 0.

The lensing amplification factor $F(f) = \tilde{h}^L(f)/\tilde{h}(f)$ relates the lensed waveform $\tilde{h}^L(f)$ to the unlensed waveform $\tilde{h}(f)$ for GWs of frequency $f$.  It is given by Kirchhoff's diffraction integral \cite{1992grle.book.....S,Takahashi_2003}
\begin{align}\label{eq:amplification factor}
    F(f) = \frac{D_S \xi^2_0 (1+z_L)}{D_L D_{LS}} \frac{f}{i} \int d^2 \vec{x}~ \text{exp}{[2 \pi i f t_d(\boldsymbol{x}, \boldsymbol{y})]}~.
\end{align}
This integral over the lens plane accounts for all the trajectories in which the wave can propagate; it is unity in the absence of a lens.

\subsubsection{\label{subsec: geometrical optics approximation} Geometrical-optics approximation}

In the geometrical-optics approximation, generally valid for GW frequencies $f$ and lens masses $M_L$ for which $f \gg M_L^{-1}$ \cite{Takahashi_2004}, discrete images form at the stationary points $\vec{x}_j$ of the time-delay function at which $\nabla_x t_d(\vec{x}, \vec{y}) = 0$.  Only these points contribute to the lensing amplification factor
\begin{align}\label{eq:amplification factor geo}
    F(f) = \sum_j |\mu_j|^{1/2} \text{exp}\left(2 \pi i f t_d({\vec{x}_j, \vec{y}}) - i \pi n_j \right)\,,
\end{align}
where $\mu_j = 1/{\rm det}(\partial\vec{y}/\partial\vec{x}_j)$ is the magnification of the $j^{th}$ image and the Morse index $n_j$ has values of 0, 1/2, or 1 depending on whether $\vec{x}_j$ is a minimum, saddle point, or maximum respectively of the time-delay surface, $t_d(\vec{x}, \vec{y})$. 

\subsection{\label{subsec:gravitational waveform} Gravitational waveform}

We restrict our analysis to the inspiral phase of the GW evolution from binary black hole (BBH) mergers and use the post-Newtonian (PN) approximation to model our unlensed waveform \cite{PhysRevD.49.2658}
\begin{align} \label{eq:gw waveform}
    \tilde{h}(f)= 
    \begin{cases}
    \frac{A}{D} \mathcal{M}^{5/6} f^{-7/6} e^{i \Psi(f)},&  0 < f < f_{\rm cut}\\
    0,              & f_{\rm cut} < f\,,
    \end{cases}
\end{align}
where $D$ is the the luminosity distance to the source, $\Psi(f)$ is the GW phase, and the GW amplitude $A$ is a function of sky localization and source geometry of order unity as discussed in \cite{1994PhRvD..49.6274A}. For a BBH system with masses $m_1$ and $m_2$, $M = m_1 + m_2$ is the total mass, $\eta = m_1m_2/M^2$ is the symmetric mass ratio, $M_z = (1+z)M$ is the redshifted total mass, and $\mathcal{M} = \eta^{3/5}M_z$ is the redshifted chirp mass. The cutoff frequency $f_{\rm cut} = 1/(6^{3/2} \pi M_z)$ is chosen to be twice the orbital frequency at the innermost stable circular orbit of a BH of mass $M_z$.

To 1.5PN order \cite{PhysRevD.49.2658}, the GW phase is 
\begin{align} \label{eq:gw phase}
    \Psi(f) &= 2\pi f t_c - \phi_c - \frac{\pi}{4} \notag \\
    & \quad + \frac{3}{4} (8 \pi \mathcal{M} f)^{-\frac{5}{3}} \left[1 + \frac{20}{9} \left(\frac{743}{336} + \frac{11 \eta}{4}\right) x - 16 \pi x^\frac{3}{2}\right],
\end{align}
where $t_c$ and $\phi_c$ are the coalescence time and phase and $x \equiv (\pi M_z f)^{2/3}$ is the PN expansion parameter.

\section{\label{sec:axisymmetric lens}Axisymmetric lens models}

In this section, we discuss two axisymmetric lens models, the singular isothermal sphere (SIS) and the point mass (PM), that produce at most two images in the geometrical-optics approximation.  We introduce model-independent image parameters that describe the amplification factor $F$ in this approximation, and assess the validity of these new parameters as the geometrical-optics approximation breaks down at low frequencies.

\subsection{\label{subsec:Singular Isothermal Sphere lens}Singular isothermal sphere (SIS)}

The SIS density profile $\Sigma(\boldsymbol{\xi}) = \sigma_v^2/2\xi$, where $\sigma_v$ is the velocity dispersion, is the most simple profile that can effectively describe the flat rotation curves of galaxies \cite{Gavazzi_2007}. It leads to the lensing potential $\psi(x) = x$ by Eq.~(\ref{E:Poisson}) and the amplification factor \cite{Takahashi_2003,Matsunaga_2006}
\begin{align}
    F(f) &= -iw\text{e}^{iwy^2/2} \int_0^\infty dx\,x J_0(wxy) \nonumber \\
    & \quad \quad \times \exp \left[iw \left(\frac{1}{2}x^2 -x +\phi_m(y) \right) \right]
    \nonumber \\
    &= e^{\frac{i}{2}w(y^2 + 2\phi_m(y))} \sum_{n=0}^\infty \frac{\Gamma(1 + \frac{n}{2})}{n!} \nonumber \\
    & \quad \quad \times (2we^{i\frac{3 \pi}{2}})^\frac{n}{2} {}_{1}F_1 \left(1 + \frac{n}{2}, 1; -\frac{i}{2}w y^2\right)\,, \label{eq: amp factor sis analytical}
\end{align}
by Eq.~(\ref{eq:amplification factor}), where $w = 8\pi M_L f$, $\phi_m(y) = y + 1/2$, $J_0$ is the Bessel function of zeroth order, $\xi_0 = 4\pi^2\sigma_v^2D_{L}D_{LS}/D_S$ is the Einstein radius, and $M_L = \sigma_v^2(1+z_L)\xi_0$ is the lens mass inside the Einstein radius.

\subsection{\label{subsec: Point mass lens}Point mass (PM)}

The PM is the simplest mass distribution for a gravitational lens. It leads to a lensing potential $\psi(x) = \ln x$ and amplification factor \cite{PhysRevD.9.2207, Takahashi_2003}
\begin{align}\label{eq: amp factor pm}
    F(f) &= \text{exp} \left\{ \frac{\pi w}{4} + i \frac{w}{2} \left[\ln\left(\frac{w}{2}\right) - 2 \phi_m(y) \right]  \right\} \nonumber \\& \times  \Gamma \left(1 - \frac{i}{2}w \right) {}_{1}F_1 \left(\frac{i}{2}w, 1; \frac{i}{2}w y^2\right)\,,
\end{align}
where $w = 8 \pi M_L f$, $\phi_m(y) = (x_m - y)^2/2 - \ln x_m$, $x_m = (y + (y^2 + 4)^{1/2})/2$, and ${}_{1}F_1(a, b; c)$ is the confluent hypergeometric function. The Einstein radius $\xi_0$ for a PM of lens mass $M_L$ is $\xi_0 = [4M_LD_L D_{LS}/D_S(1 + z_L)]^{1/2}$.

\begin{figure}[t!]
    \centering
    \includegraphics[scale = 0.33]{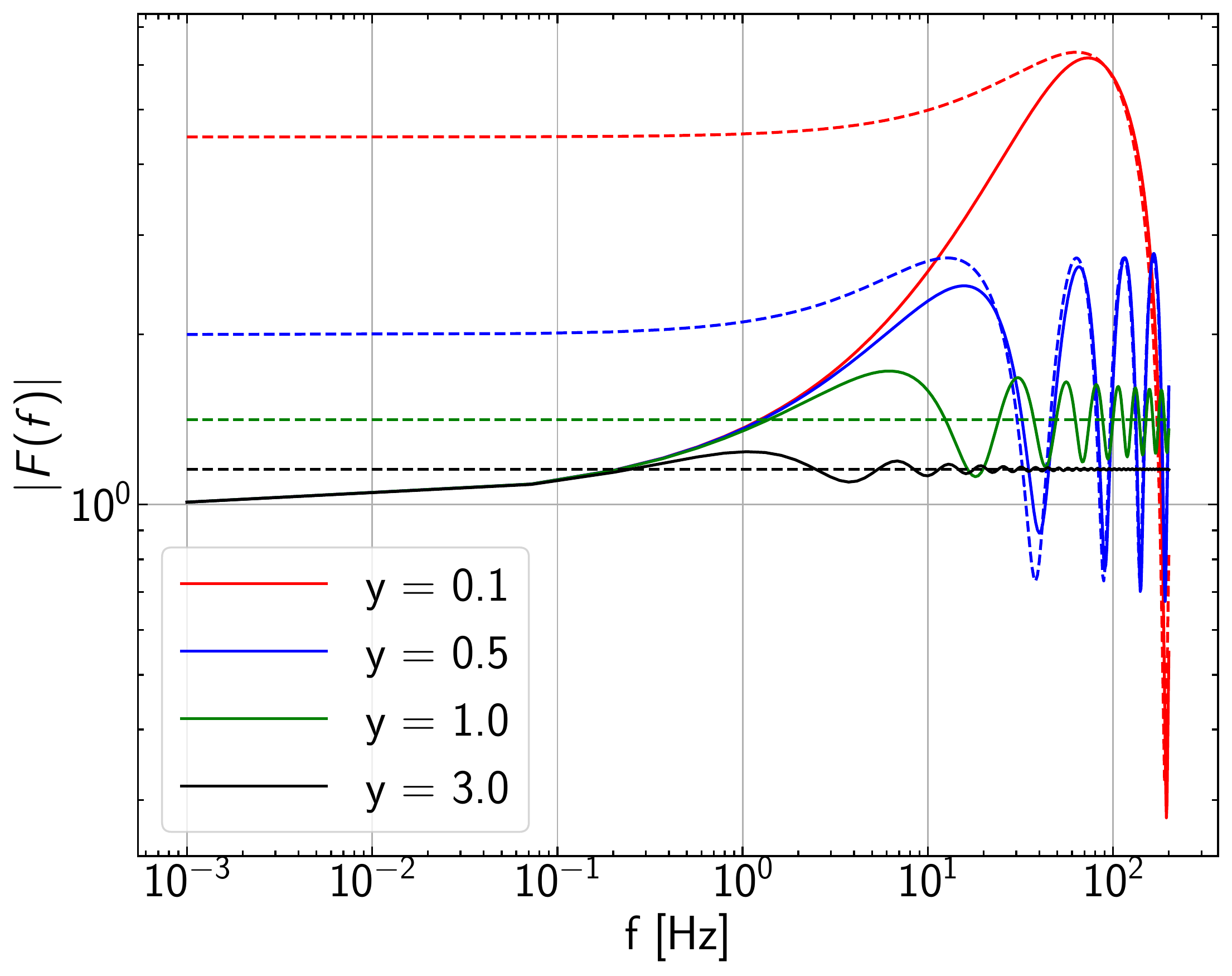}
    \caption{The magnitude of the amplification factor $F(f)$ as a function of frequency $f$ for an SIS with lens mass $M_L = 10^3 M_\odot$ and source positions $y = 0.1$, $0.5$, $1.0$, and $3.0$ shown in red, blue, green, and black. The solid lines are calculated using the exact result of Eq.~(\ref{eq: amp factor sis analytical}), while the dashed lines are calculated using the geometrical-optics approximation of Eq.~(\ref{eq:amp factor sis geo}).
    }
    \label{fig:amplification factor sis}
\end{figure}

\subsection{\label{subsubsec:geometrical optics approximation} Geometrical-optics approximation}

In the limit where the geometrical-optics approximation is valid, a source at position $\vec{y}$ creates a discrete number of images at positions $\vec{x}_j$ which are stationary points of the time delay $t_d(\vec{x}, \vec{y})$ given by Eq.~(\ref{eq:time delay}). For the SIS lens, only one image is formed if the source is outside the unit circle ($y > 1$), whereas two images are formed if the source is inside ($y < 1$). The amplification factor $F(f)$ in the geometrical-optics approximation of  Eq.~(\ref{eq:amplification factor geo}) is given by \cite{Takahashi_2003}
\begin{align}\label{eq:amp factor sis geo}
    F(f) = 
    \begin{cases}
    |\mu_+|^{1/2} - i |\mu_-|^{1/2} e^{2 \pi i f \Delta t_d},&  y < 1\\
    |\mu_+|^{1/2},              & y > 1\,,
    \end{cases}
\end{align}
where
\begin{subequations} \label{E:SISip}
\begin{align}
    \mu_\pm &= \pm 1 + \frac{1}{y}\,, \label{E:SISmag} \\
    \Delta t_d &= 8M_Ly\,. \label{E:SIStd}
\end{align}
\end{subequations}
As a PM lens can deflect photons by an arbitrarily large angle for sufficiently small impact parameter, there are two images for all source positions $y$. $F(f)$ is given by \cite{Takahashi_2003}
\begin{equation}\label{eq:amp factor pm geo}
    F(f) = |\mu_+|^{1/2} - i |\mu_-|^{1/2} e^{2 \pi i f \Delta t_d}\,,
\end{equation}
where
\begin{subequations} \label{E:PMip}
\begin{align}
    \mu_\pm &= \frac{1}{2} \pm \frac{y^2 + 2}{2y(y^2 + 4)^{1/2}}\,,
    \label{E:PMmag} \\
    \Delta t_d &= 2M_L \left\{ y(y^2 + 4)^{1/2} + 2\ln \left[ \frac{(y^2 + 4)^{1/2} + y)}{(y^2 + 4)^{1/2} - y)} \right] \right\}\,. \label{E:PMtd}
\end{align}
\end{subequations}
Both the SIS and PM lens models can be parameterized by two lens parameters, the source position $y$ and lens mass $M_L$. 

The amplification factor $F(f)$ for the SIS is shown in Fig.~\ref{fig:amplification factor sis}. In the low-frequency limit, $F(f)$ converges to unity because diffraction effects prevent long-wavelength GWs from being affected by the presence of a lens \cite{Bontz1981ADL, Takahashi_2003}. In the high-frequency limit, the geometrical-optics approximation is valid and oscillatory behavior is observed due to interference between the coherent lensed images. We assume that the time delay $\Delta t_d$ is much less than the observing time of the GW detector as is appropriate for the lens mass shown in Fig.~\ref{fig:amplification factor sis}.
\begin{figure}[t!]
    \centering
    \includegraphics[scale = 0.33]{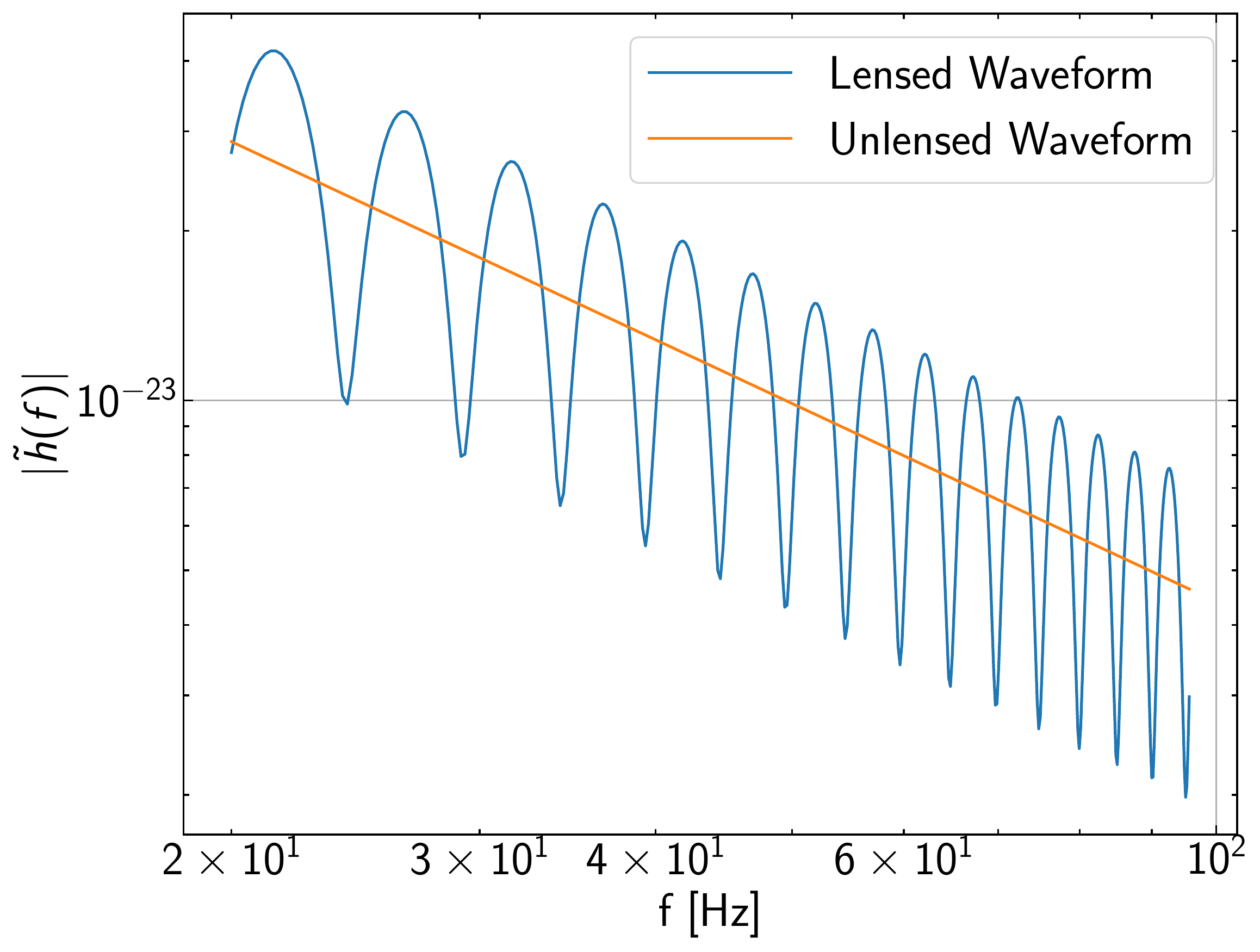}
    \caption{Waveform magnitudes as functions of frequency $f$ for our default source parameters of $A = 0.21$, $\mathcal{M} = 20 M_\odot$, $\eta = 0.25$, $t_c = \phi_c = 0$, and $D = 1\,$Gpc. The orange line shows the unlensed waveform $\tilde{h}(f)$, while the blue curve shows the lensed waveform $\tilde{h}^L(f) = F(f)\tilde{h}(f)$ for the amplification factor $F(f)$ of a SIS with lens parameters $y = 0.5$ and $M_L = 10^4 M_\odot$ calculated in the geometrical-optics approximation.
    }
    \label{fig:sis lensed and unlensed waveforms}
\end{figure}

Fig.~\ref{fig:sis lensed and unlensed waveforms} shows the unlensed waveform $\tilde{h}(f)$ and lensed waveform $\tilde{h}^L(f) = F(f)\tilde{h}(f)$ as functions of frequency $f$. The unlensed waveform given by Eq.~(\ref{eq:gw waveform}) is parameterized by the GW amplitude $A$, redshifted chirp mass $\mathcal{M}$, symmetric mass ratio $\eta$, source luminosity distance $D$, coalescence time $t_c$, and coalescence phase $\phi_c$; we choose fiducial values of $A = 0.21$, $\mathcal{M} = 20 M_\odot$, $\eta = 0.25$, $t_c = \phi_c = 0$, and $D = 1\,$Gpc for these source parameters. It is proportional to $f^{-7/6}$ according to Eq.~(\ref{eq:gw waveform}) and appears as a straight line in this log-log plot.
The lensed waveform is calculated for a SIS with lens parameters $y = 0.5$ and $M_L = 10^4 M_\odot$ and displays oscillatory behavior due to interference between the two terms in Eq.~(\ref{eq:amp factor sis geo}).

\subsection{\label{subsec: model-independent lens parameters}Model-independent image parameters}

When the amplification factor $F(f)$ is given by Eq.~(\ref{eq:amp factor pm geo}), i.e. the source position is such that two images are formed and the geometrical-optics approximation is valid, we can express this factor directly in terms of the flux ratio $I = |\mu_-|/|\mu_+|$ and time delay $\Delta t_d$ between the images. These "image parameters" are model-independent in that they fully specify the amplification factor for any two-image lens model up to an overall normalization that is observationally degenerate with the GW amplitude $A$.  The relationship between these image parameters and lens parameters like the source position $y$ and lens mass $M_L$ is model dependent; these relations for the SIS and PM lens models are given by Eqs.~(\ref{E:SISip}) and (\ref{E:PMip}) respectively.
\begin{figure}[t!]
    \centering
    \includegraphics[scale = 0.27]{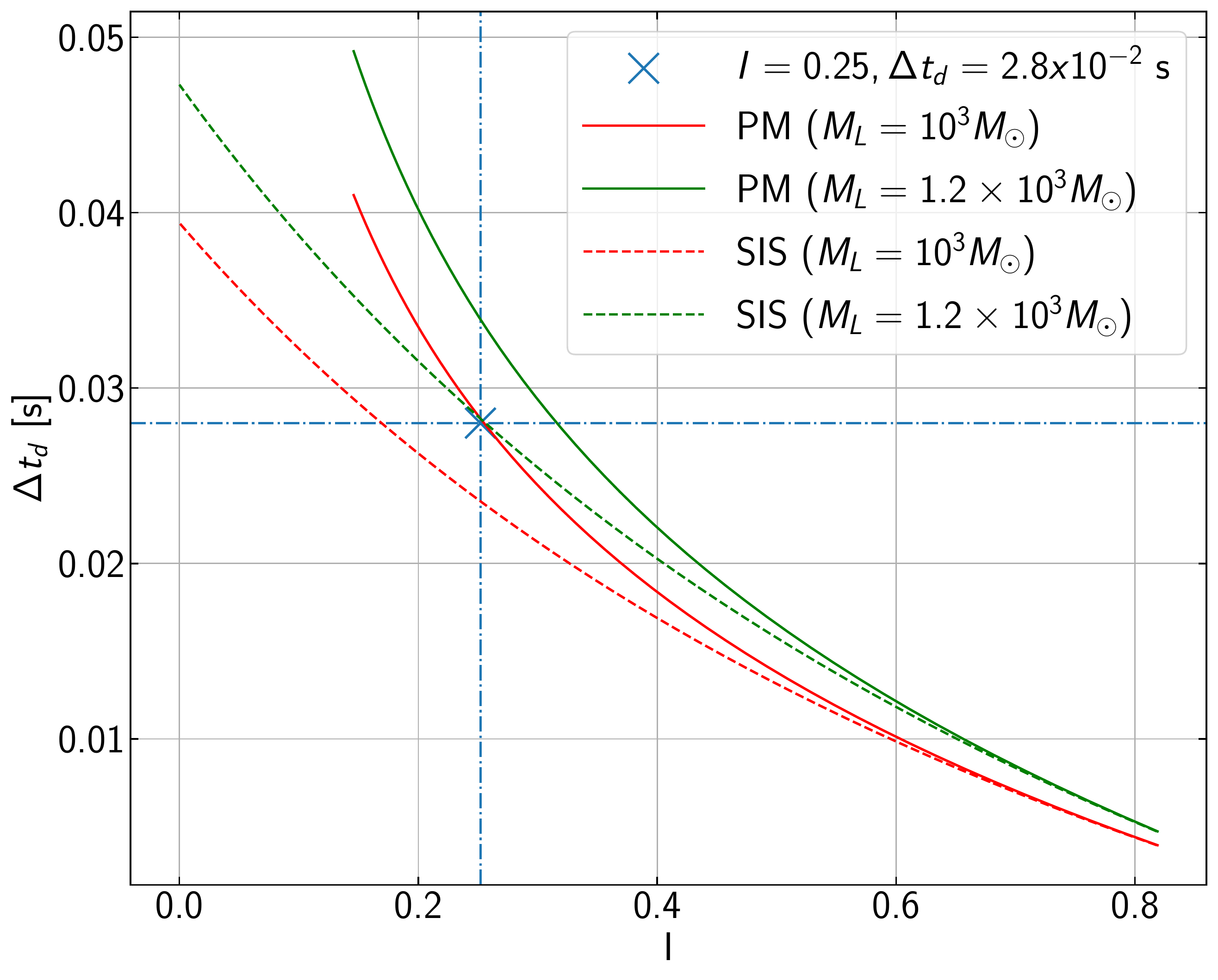}
    \caption{
    The flux ratio $I$ and time delay $\Delta t_d$ as parametric functions of the source position $y$ for two-image lens models of varying lens mass $M_L$.  The solid (dashed) curves correspond to the PM (SIS) lens model, while the red (green) curves are for a lens mass $M_L = 10^3 M_\odot$ ($1.2 \times 10^3 M_\odot$). The blue cross indicates the point ($I = 0.25,\,\Delta t_d = 2.8 \times 10^{-2}$ s) where curves for the PM and SIS models intersect.
    }
    \label{fig:I-td sis}
\end{figure}

In Fig.~\ref{fig:I-td sis}, we plot the flux ratio $I$ and time delay $\Delta t_d$ as parametric functions of the source position $y$ for our two axisymmetric lens models and two choices of the lens mass $M_L$. The blue cross indicates specific, potentially observable values ($I = 0.25$ and $\Delta t_d = 2.8 \times 10^{-2}$ s) of the model-independent image parameters. According to Eqs.~(\ref{E:SISip}) and (\ref{E:PMip}), these values can be obtained in both the SIS and PM lens models, albeit with different lens parameters, $M_L \approx 1.2 \times 10^3 M_\odot$ and $y \approx 0.6$ for the SIS lens and $M_L \approx 10^3 M_\odot$ and $y \approx 0.7$ for the PM lens.  This analysis reveals that apart from any observational errors associated with measuring the image parameters $I$ and $\Delta t_d$, there is a $\sim20\%$ model dependence with which lens parameters like $M_L$ and $y$ can be reconstructed.
\begin{figure*}[t!]
    \centering
    \includegraphics[scale = 0.29]{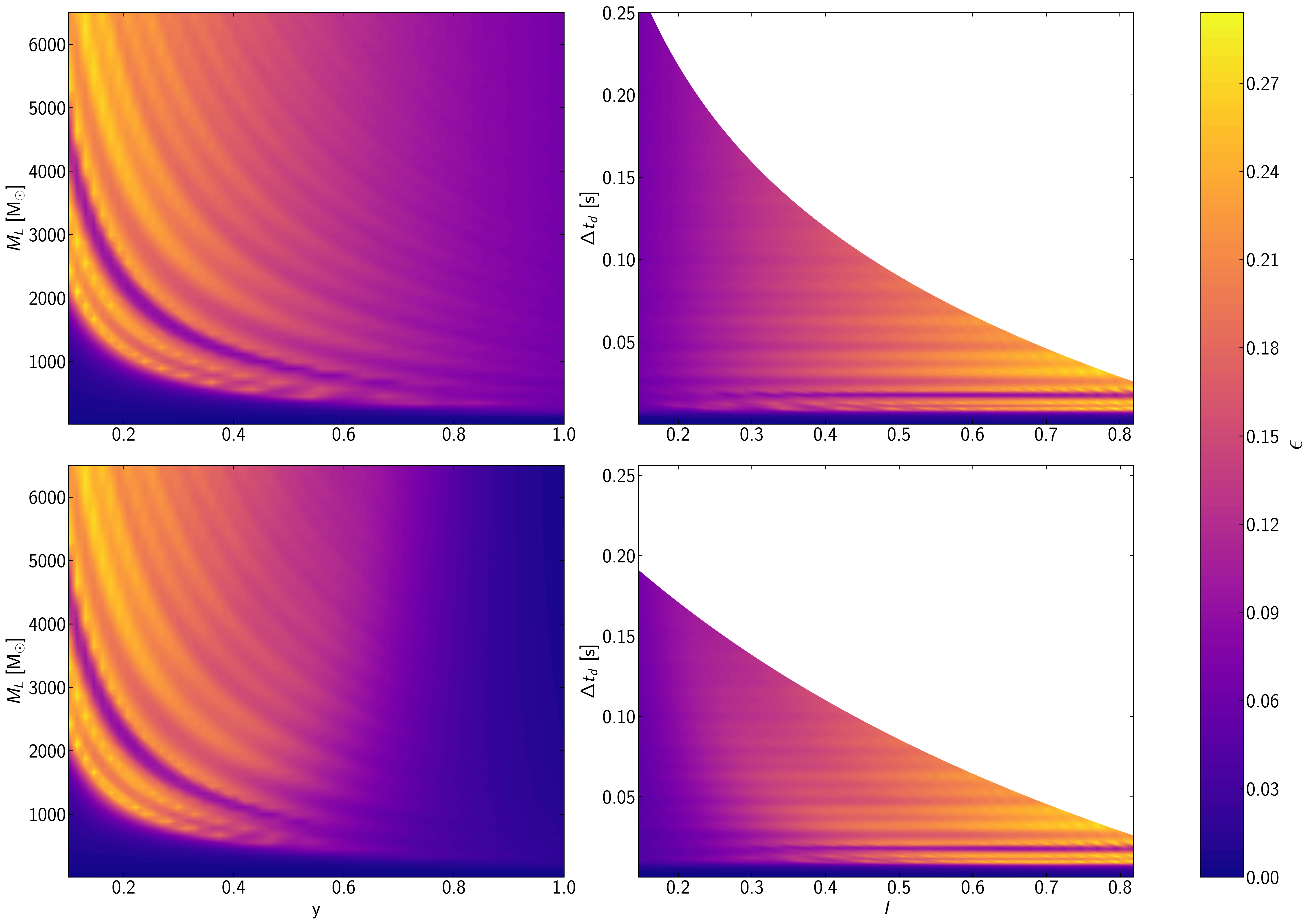}
    \caption{
    Contour plots of the mismatch $\epsilon$ between lensed and unlensed GWs for PM lenses (top panels) and SIS lenses (bottom panels). The left panels show $\epsilon$ as a function of the model-dependent source distance $y$ and lens mass $M_L$, while the right panels show $\epsilon$ as a function of the model-independent flux ratio $I$ and time delay $\Delta t_d$. The unlensed waveforms are calculated using the default source parameters $\mathcal{M} = 20 M_\odot$, $\eta = 0.25$, and $t_c = \phi_c = 0$.
    } 
    \label{fig:contour mismatch y-ML}
\end{figure*}

\section{\label{sec: Match filtering analysis} Matched-filtering analysis}

In this section, we perform a matched-filtering analysis to quantify the difference between lensed and unlensed GWs. The mismatch $\epsilon$ between two waveforms $h_1$ and $h_2$ is defined as \cite{PhysRevD.49.2658}
\begin{equation}\label{eq:mismatch}
    \epsilon (h_1, h_2) \equiv 1 -  \max\limits_{t_c, \phi_c} \frac{<h_1|h_2>}{\sqrt{<h_1|h_1><h_2|h_2>}}\,.
\end{equation}
The noise-weighted inner product $<h_1|h_2>$ between the waveforms $h_1$ and $h_2$ is defined as
\begin{equation}\label{eq:inner product}
    <h_1|h_2> = 4\,\text{Re} \int_{f_{\rm low}}^{f_{\rm cut}} df \frac{h_1(f) h_2^*(f)}{S_n(f)}\,,
\end{equation}
where $S_n(f)$ is the noise power spectral density (PSD). We use the \textsc{pycbc.filter} package \cite{alex_nitz_2022_6324278} to evaluate the mismatch between waveforms. The two waveforms $h_1$ and $h_2$ can be distinguished when their mismatch $\epsilon \gtrsim \rho^{-2}$, where $\rho = <h|h>^{1/2}$ is the signal-to-noise ratio (SNR) of a waveform $h$ \cite{1992PhRvD..46.5236F,1993PhRvD..47.2198F,PhysRevD.49.2658}.

Fig.~\ref{fig:contour mismatch y-ML} shows contour plots of the mismatch $\epsilon$ between lensed GW source and unlensed templates as functions of the model-dependent lens parameters $y$ and $M_L$ (left panels) and the model-independent image parameters $I$ and $\Delta t_d$ (right panels) for PM lenses (top panels) and SIS lenses (bottom panels). We use the same source parameters for the lensed and unlensed templates and the noise PSD appropriate for a single two-armed detector with aLIGO design sensitivity.  The mismatches as functions of the lens parameters are qualitatively similar for the PM (top left panel) and SIS (bottom left panel) models.  For a vanishing lens mass ($M_L \to 0$), diffraction causes the amplification factor to approach unity ($F \to 1$) and the mismatch to vanish ($\epsilon \to 0$) according to Eq.~(\ref{eq:mismatch}) for $h_1 = h_2$.  The biggest difference between the two models occurs in the limit $y \to 1$, where $|\mu_-| \to 0$ for the SIS model according to Eq.~(\ref{E:SISmag}) but $|\mu_-| \to 0.17$ for the PM model according to Eq.~(\ref{E:PMmag}).  This accounts for the much smaller mismatches for the SIS model compared to the PM model in this limit.

The PM and SIS models become qualitatively indistinguishable when the mismatch is expressed as a function of the image parameters of $I$ and $\Delta t_d$ as shown in the right panels of Fig.~\ref{fig:contour mismatch y-ML}.  Although the top and bottom panels appear the same in the regions where they overlap, the PM model extends to larger values of the time delay $\Delta t_d$ than the SIS model at small flux ratios $I$.  This occurs because the mappings Eqs.~(\ref{E:SISip}) and (\ref{E:PMip}) between the lens parameter space in the left panels and the image parameter space in the right panels are model-dependent.

\begin{figure*}[ht!]
    \centering
    \includegraphics[scale = 0.43]{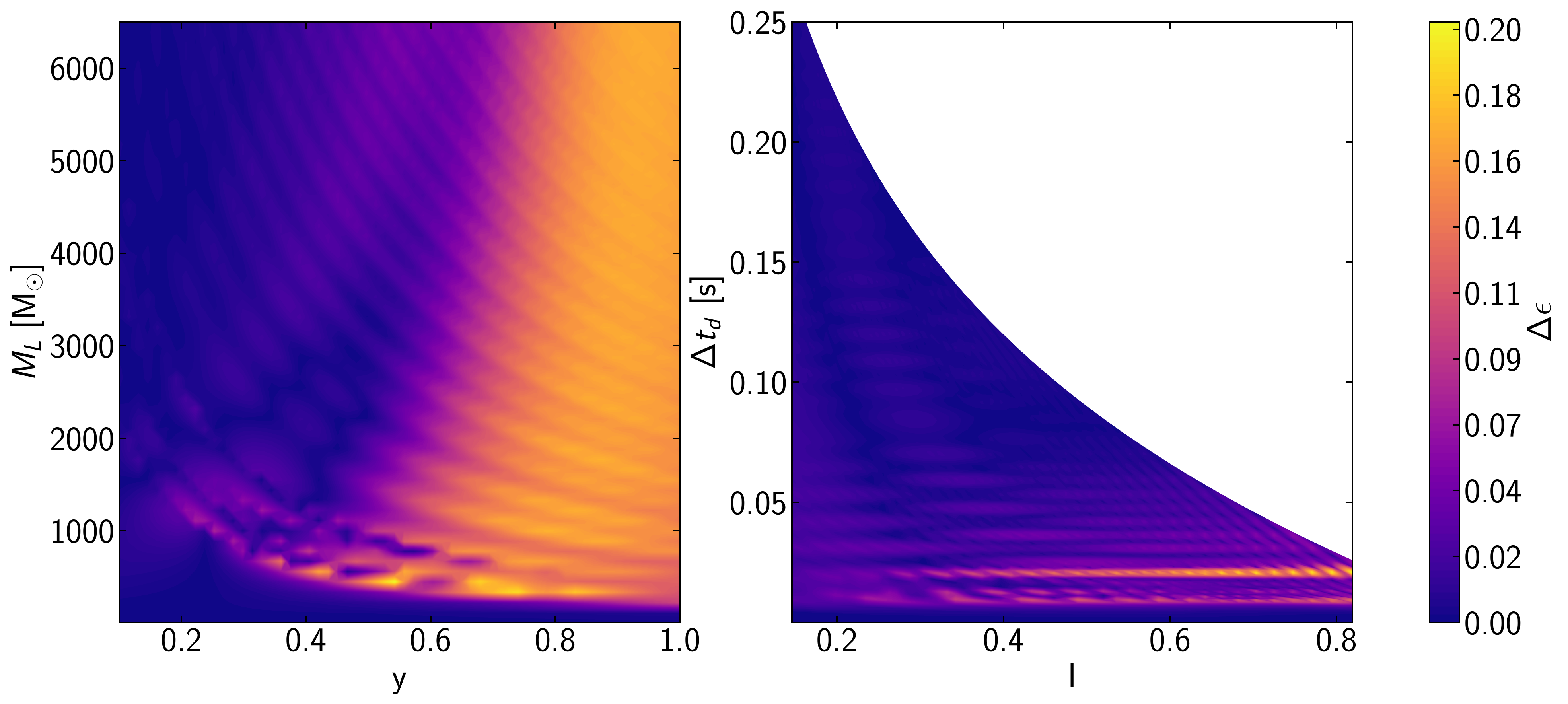}
    \caption{
    The difference $\Delta\epsilon = |\epsilon_{\rm PM} - \epsilon_{\rm SIS}|$ between the mismatches between lensed and unlensed waveforms for PM and SIS lens models shown in the top and bottom panels of Fig.~\ref{fig:contour mismatch y-ML} as functions of the source position $y$ and lens mass $M_L$ (left panel) and the flux ratio $I$ and time delay $\Delta t_d$ (right panel).
    } 
    \label{fig:contour mismatch compare}
\end{figure*}

The better agreement between the mismatches in the PM and SIS lens models when expressed as functions of the image parameters $I$ and $\Delta t_d$ can be seen even more clearly in Fig~\ref{fig:contour mismatch compare}.  This figure shows the difference $\Delta\epsilon = |\epsilon_{\rm PM}-\epsilon_{\rm SIS}|$ between the mismatches for PM and SIS lens models shown in the top and bottom panels of Fig.~\ref{fig:contour mismatch y-ML}.  The left panel of Fig~\ref{fig:contour mismatch compare} shows large differences $\Delta\epsilon \gtrsim 0.15$ between the two lens models for source distances $y \gtrsim 0.7$ where the flux ratio $I$ vanishes in the SIS model but not the PM model.  However, when the mismatch difference is expressed as a function of the image parameters in the right panel, the amplification factors and thus the mismatches only have significant differences for $\Delta t_d \lesssim 0.02$s where the geometrical-optics approximation breaks down. 

\begin{figure*}
    \centering
    \includegraphics[scale = 0.28]{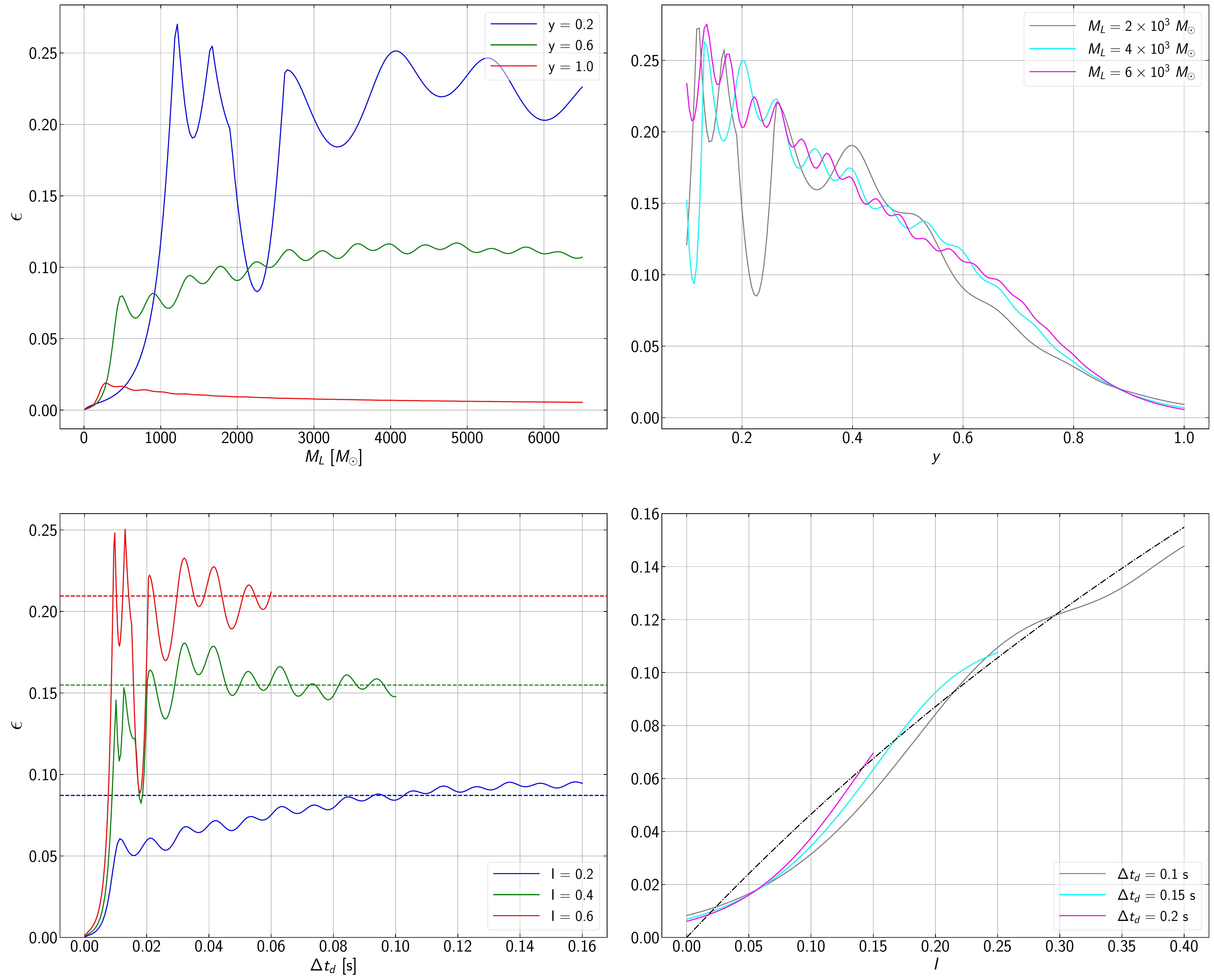}
    \caption{
    The mismatch $\epsilon$ between lensed and unlensed waveforms with source parameters $\mathcal{M} = 20 M_\odot$, $\eta = 0.25$, and $t_c = \phi_c = 0$ in the SIS lens model as a function of lens mass $M_L$ (top left panel), source position $y$ (top right panel), time delay $\Delta t_d$ (bottom left panel), and flux ratio $I$ (bottom right panel). In the top (bottom) left panel, the blue, green, and red curves correspond to $y$ = 0.2, 0.6, and 1.0 ($I$ = 0.2, 0.4, and 0.6).  In the top (bottom) right panel, the grey, cyan, and magenta curves correspond to $M_L = 2 \times 10^3 M_\odot$, $4 \times 10^3 M_\odot$, and $6 \times 10^3 M_\odot$ ($\Delta t_d$ = 0.1 s, 0.15 s, and 0.2 s).  The dotted curves in the bottom panels show the mismatch $\epsilon \approx 1 - (1 + I)^{-1/2}$ in the extreme geometrical-optics limit derived in Appendix~\ref{appendix:Mismatch approximation in geometrical-optics limit}.
    }
    \label{fig:1d mismatch y-ML}    
\end{figure*}

The right panels of Fig.~\ref{fig:contour mismatch y-ML} also reveal that the crests and troughs of the oscillations in $\epsilon$ occur on lines of constant time delay $\Delta t_d$.  We examine these oscillations and the general dependence of the mismatch on our lens and image parameters for the SIS model in Fig.~\ref{fig:1d mismatch y-ML}.  The top left (right) panels of Fig.~\ref{fig:1d mismatch y-ML} show vertical (horizontal) slices of the bottom left panel of Fig.~\ref{fig:contour mismatch y-ML} along lines of constant $y$ ($M_L$).  The dependence of the mismatch $\epsilon$ on these lens parameters is non-monotonic and difficult to interpret, exhibiting oscillations of varying amplitude, frequency, and phase and a deep valley.  The dependence of $\epsilon$ on the image parameters $\Delta t_d$ and $I$ shown in the bottom panels is far easier to interpret.  The bottom left panel shows that $\epsilon \to 0$ in the diffraction limit $\Delta t_d \to 0$, while in the opposite extreme geometrical-optics limit $\Delta t_d \to \infty$, $\epsilon \to 1 - (1 + I)^{-1/2}$ as derived in Appendix~\ref{appendix:Mismatch approximation in geometrical-optics limit}.  The crests and troughs of the oscillations, as well as the deep valley at $\Delta t_d \approx 0.015$\,s are all aligned in this panel, reinforcing our contention that these oscillations are purely functions of the time delay $\Delta t_d$.  They are absent in the bottom right panel, where $\Delta t_d$ is held constant and the mismatch depends smoothly on the flux ratio $I$ in reasonable agreement with the extreme geometrical-optics limit $\epsilon \to 1 - (1 + I)^{-1/2}$.

We investigate the oscillations in the mismatch $\epsilon$ as a function of the time delay $\Delta t_d$ in Appendix~\ref{appendix:Investigating striations in the mismatch}.  These oscillations result from the number and location of the peaks of the amplification factor $F(f)$ shown in Fig.~\ref{fig:amplification factor sis} within the sensitivity band of the detector changing as  $\Delta t_d$ varies.  For our approximate waveform of Eq.~(\ref{eq:gw waveform}), the sharpest feature in the detector response is a cutoff in the strain $h(f)$ above a frequency $f_{\rm cut} = (6^{3/2}\pi M_z)^{-1}$.  According to Eq.~(\ref{E:depdtd}), this cutoff couples to the peaks of the amplification factor to create oscillations in the mismatch with frequency as a function of the time delay with frequency $f_{\rm cut}$ and amplitude proportional to $I^{3/2}/(1+I)$.  Numerical-relativity simulations reveal that true waveforms transition smoothly to a ringdown rather than experiencing such a sharp cutoff \cite{Mroue:2013xna, LIGOScientific:2014oec, Boyle:2019kee}, but the merger should still imprint a feature leading to oscillations in the mismatch between a lensed GW source and unlensed templates.  Our analysis suggests that these oscillations may be significant for large flux ratios $I$ (small source distances $y$), implying that serendipity between the lens parameters and source mass may facilitate the discovery of lensing in GW events.

\section{\label{sec:Conclusion}Discussion}

In the geometrical-optics approximation, strong gravitational lensing produces multiple images. The lensing amplification factor $F(f)$ can be expressed as a summation over these images, each with its own magnification $\mu_j$, time delay $t_{d,j}$, and Morse index $n_j$ as given by Eq.~(\ref{eq:amplification factor geo}). In any given lens model, these "image parameters" can be calculated as functions of the "lens parameters" such as the source position $\vec{y}$ and the lens mass $M_L$.  We propose that observational searches for strong lensing in GW events, as well as theoretical studies of their feasibility, should be conducted in terms of these model-independent image parameters rather than the model-dependent lens parameters.  Separating detection and parameter estimation/model selection into distinct stages of GW data analysis should clarify the challenges of each stage as well as reduce their associated computational costs.

In this paper, we investigate this proposal by considering two well-known axisymmetric lens models: the singular isothermal sphere (SIS) and the point mass (PM). The SIS model has a density profile $\rho \propto r^{-2}$ appropriate for a galaxy or stellar cluster, while the PM model could describe an individual star or compact object.  In the geometrical-optics approximation, the SIS produces two images for source positions $y < 1$, while the PM always produces two images. For the case of two images, the amplification factor is given by Eq.~(\ref{eq:amp factor pm geo}) for both models, but the mappings between the image parameters (the flux ratio $I = |\mu_-|/|\mu_+|$ and the time delay $\Delta t_d$) and the lens parameters (the source distance $y$ and the lens mass $M_L$) are model dependent and given by Eqs.~(\ref{E:SISip}) and (\ref{E:PMip}) respectively.  This implies that uncertainty in the lens model will translate into uncertainty in the lens parameters, even if the image parameters could be measured with arbitrary precision.  We illustrate this in Fig.~\ref{fig:I-td sis}, where the SIS and PM models can generate images with identical flux ratios and time delays despite have lens parameters that differ by $\sim20\%$.  This represents a conservative estimate of the systematic error in the lens parameters associated with uncertainty in the lens model, as the PM model is more compact than any real galaxy or halo.

Lensed GW events with finite signal-to-noise ratios $\rho$ can be distinguished from unlensed templates if the minimum mismatch $\epsilon$ between the lensed signal and members of the template bank exceeds $\rho^{-2}$ \cite{1992PhRvD..46.5236F,1993PhRvD..47.2198F,PhysRevD.49.2658}.  The oscillatory features induced in the waveform by lensing in the geometrical-optics limit depicted in Fig.~\ref{fig:sis lensed and unlensed waveforms} have little degeneracy with the source parameters specifying the unlensed waveform in Eqs.~(\ref{eq:gw waveform}) and (\ref{eq:gw phase}), so we approximate this minimum mismatch by that between lensed and unlensed waveforms with the same source parameters.  In this approximation, the mismatch can be approximated by $\epsilon \approx 1 - (1 + I)^{-1/2}$ according to Eq.~(\ref{E:MMexgeo}) and lensing should in principle be identifiable for flux ratios $I \gtrsim 2\rho^{-2}$ and time delays $\Delta t_d \gtrsim f_{\rm cut}^{-1}$.  We further showed in the right panel of Fig.~\ref{fig:contour mismatch compare} that the amplification factor of Eq.~(\ref{eq:amp factor pm geo}) is an excellent approximation to both the SIS and PM lens models in the geometrical-optics limit, suggesting that it should be suitable for model-independent searches for strong lensing in GW events.

Although gravitational lensing has not been detected in any of the GW events observed in the first three runs of the LVK detector network \cite{abbott2021search}, recent estimates \cite{2018-oguri} suggest that $\sim 10^3$ strongly lensed GW events could be observed each year by future third-generation detectors like the proposed Cosmic Explorer (CE) \cite{evans2021horizon} and Einstein Telescope (ET) \cite{Maggiore:2019uih}. The detection of gravitational lensing in GW events would be of tremendous scientific interest because it could test general relativity \cite{Baker:2016reh, Collett:2016dey, Mukherjee:2019wcg}, probe the distribution of dark matter \cite{PhysRevD.106.023018, 2022A&A...659L...5C}, and improve the precision of cosmological constraints \cite{PhysRevD.103.024038} We propose that GW templates based on model-independent image parameters will be a valuable tool in this effort.  In upcoming studies, we will investigate how effectively these GW templates can be used to identify additional images created by non-axisymmetric lenses \cite{SIEinprep} and distinguish the effects of lensing from those of spin precession \cite{LensVsPrecinprep}. 

\section*{\label{sec:Acknowledgement}Acknowledgements}

This work is supported by National Science Foundation Grant No. PHY-2011977. The authors thank Christina McNally for discussions. The authors acknowledge the Texas Advanced Computing Center (TACC) at The University of Texas at Austin for providing HPC resources that have contributed to the research results reported within this paper \cite{10.1145/3093338.3093385}. URL: \href{http://www.tacc.utexas.edu }{http://www.tacc.utexas.edu}

\appendix

\section{\label{appendix:Mismatch approximation in geometrical-optics limit}Mismatch in the extreme geometrical-optics limit $\Delta t_d \to \infty$}

We define the overlap $\mathcal{O}$ between a lensed waveform $\tilde{h}^L(f)$ and unlensed template $\tilde{h}(f)$ as
\begin{align}\label{appendix eq: overlap}
    \mathcal{O} &\equiv \frac{<\tilde{h}^L(f)| \tilde{h}(f)>}{\sqrt{<\tilde{h}^L(f)|\tilde{h}^L(f)> <\tilde{h}(f)|\tilde{h}(f)>}}\,.
\end{align}
For a lensed waveform with the amplification factor $F(f)$ of Eq.~(\ref{eq:amp factor pm geo}) appropriate for a two-image lens in the geometrical-optics approximation and an unlensed waveform with the same source parameters, this overlap becomes
\begin{equation} \label{E:LensOver}
\mathcal{O} = \frac{A_1}{\sqrt{A_2A_3}},
\end{equation}
where
\begin{subequations} \label{E:OverPieces}
\begin{align}
    A_1 &\equiv \int df \frac{|\tilde{h}(f)|^2}{S_n(f)}(|\mu_+|^{1/2} + |\mu_-|^{1/2} \sin 2\pi f \Delta t_d)\,, \\
    A_2 &\equiv \int df \frac{|\tilde{h}(f)|^2}{S_n(f)}(|\mu_+| + |\mu_-| + 2 \lvert \mu_+ \mu_- \rvert^{1/2} \sin 2\pi f\Delta t_d)\,, \\
    A_3 &\equiv \int df \frac{|\tilde{h}(f)|^2}{S_n(f)}\,.
\end{align}
\end{subequations}
In the extreme geometrical-optics limit $\Delta t_d \to \infty$, the oscillatory terms do not contribute to the integrals implying $A_1 \to |\mu_+|^{1/2}A_3$, $A_2 \to (|\mu_+| + |\mu_-|)A_3$, and
\begin{equation}
\mathcal{O} \to \frac{|\mu_+|^{1/2}}{\sqrt{|\mu_+| + |\mu_-|}} = \frac{1}{\sqrt{1+I}}
\end{equation}
where $I = |\mu_-|/|\mu_+|$ is the flux ratio between the two images.
Although the match 1 - $\epsilon$ is normally calculated by maximizing the overlap  over the coalescence time $t_c$ and phase $\phi_c$ as in Eq.~(\ref{eq:mismatch}), it is independent of these parameters in the limit $\Delta t_d \to \infty$.  This implies that the mismatch takes the limiting value
\begin{equation} \label{E:MMexgeo}
\epsilon \to 1 - \mathcal{O} = 1 - \frac{1}{\sqrt{1 + I}}.
\end{equation}

\section{\label{appendix:Investigating striations in the mismatch}Investigating oscillations in the mismatch}

\begin{figure}[t!]
    \centering
    \includegraphics[scale = 0.38]{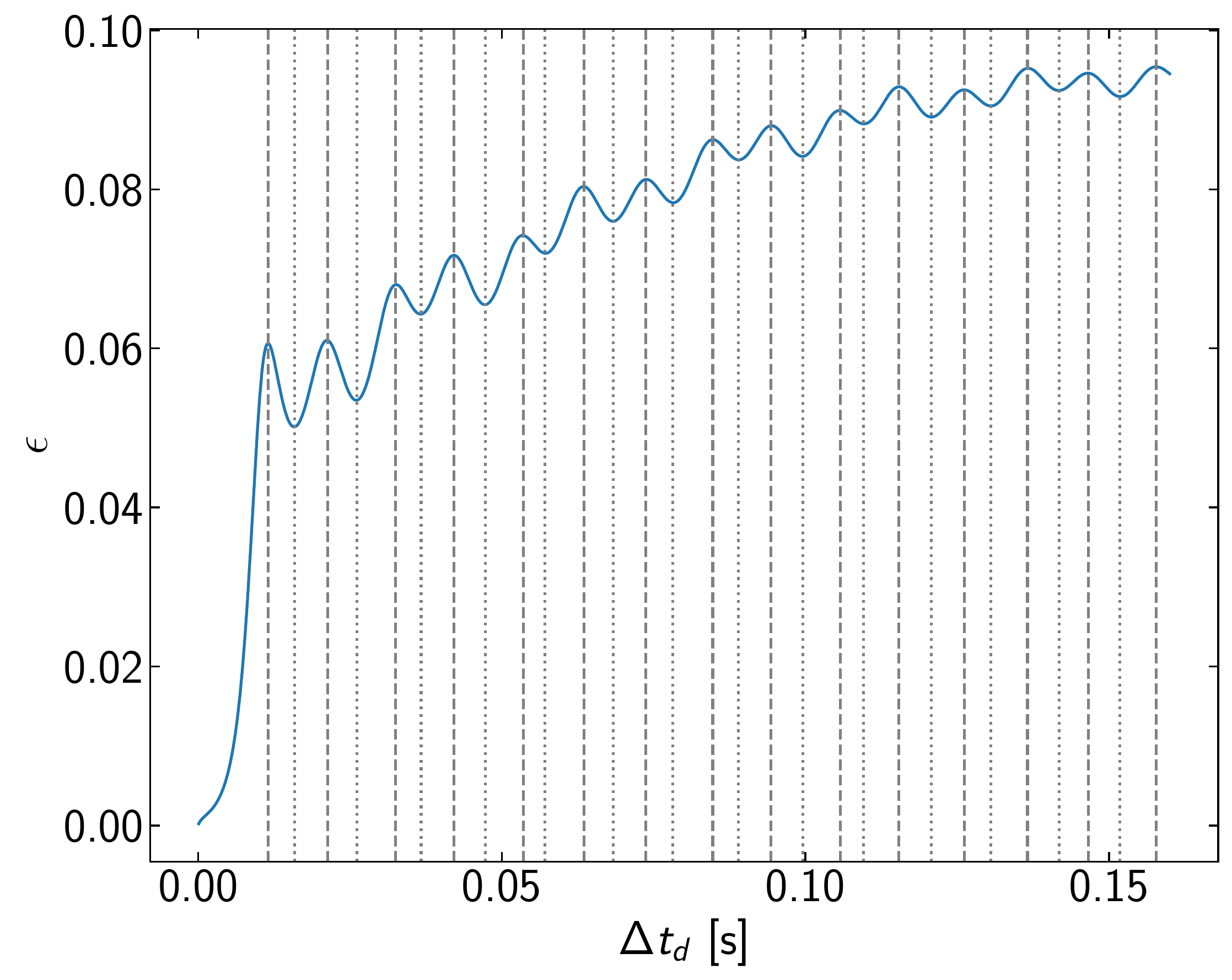}
    \caption{
    The mismatch $\epsilon$ between a lensed GW source and unlensed templates for an SIS lens as a function of time delay $\Delta t_d$ for flux ratio $I = 0.2$. The BBH system has source parameters $\mathcal{M} = 20 M_\odot$, $\eta = 0.25$, and $t_c = \phi_c = 0$.  The vertical dashed and dotted lines indicate the locations of the crests and troughs listed in Table~\ref{tab:crest_troughs_mcz_20}.
    }
    \label{fig:mismatch_mcz_20_I_0.2}
\end{figure}

In this Appendix, we examine the oscillations in the mismatch $\epsilon$ as a function of the time delay $\Delta t_d$ seen in the bottom left panel of Fig.~\ref{fig:1d mismatch y-ML}.  We reproduce the $I = 0.2$ curve from this figure in Fig.~\ref{fig:mismatch_mcz_20_I_0.2} above.  In the geometrical-optics approximation, lensing induces an oscillatory contribution to the GW phase that is poorly matched by a linear change to the GW phase resulting from a shift in the coalesence time $t_c$ and phase $\phi_c$ according to Eq.~(\ref{eq:gw phase}).  As such, the mismatch is well approximated as $\epsilon \approx 1 - \mathcal{O}$, where the overlap $\mathcal{O}$ is given by Eqs.~(\ref{appendix eq: overlap}) - (\ref{E:OverPieces}).  The crests and troughs of the oscillations in the mismatch as a function of the time delay $\Delta t_d$ will therefore occur where
\begin{align}
\frac{d\epsilon}{d\Delta t_d} &= \mathcal{O}\left(\frac{1}{2A_2}\frac{dA_2}{d\Delta t_d} - \frac{1}{A_1}\frac{dA_1}{d\Delta t_d} \right) \notag \\
&= \mathcal{O}\left( \frac{|\mu_+|^{1/2}}{A_2} - \frac{1}{A_1} \right) \frac{dA_1}{d\Delta t_d} \label{E:dOdDtd}
\end{align}
vanishes.  We show further that
\begin{subequations}
\begin{align}
\frac{dA_1}{d\Delta t_d} &= |\mu_-|^{1/2} \int_{f_{\rm low}}^{f_{\rm cut}} df \frac{|\tilde{h}(f)|^2}{S_n(f)}2\pi f\cos 2\pi f \Delta t_d \\
&\approx \frac{|\mu_-|^{1/2}}{2\pi\Delta t_d^2} \left\langle \frac{|\tilde{h}(f)|^2}{S_n(f)} \right\rangle (w\sin w + \cos w)\bigg|_{w_{\rm low}}^{w_{\rm cut}} \\
&\approx |\mu_-|^{1/2}\frac{A_3}{\Delta t_d}\sin w_{\rm cut}\,,
\end{align}
\end{subequations}
where we have defined $w \equiv 2\pi f\Delta t_d$, approximated $|\tilde{h}(f)|^2/S_n(f)$ by a constant in angular brackets equal to its average value, and assumed $w_{\rm cut} \gg w_{\rm low} \gg 1$ as will be true for small redshifted source masses in the geometrical-optics limit.  Inserting these into Eq.~(\ref{E:dOdDtd}) and using the extreme geometrical-optics approximation described in Appendix~\ref{appendix:Mismatch approximation in geometrical-optics limit}, we find
\begin{equation} \label{E:depdtd}
\frac{d\epsilon}{d\Delta t_d} = -\frac{I^{3/2}}{1 + I} \frac{\mathcal{O}}{\Delta t_d} \sin w_{\rm cut}\,.
\end{equation}
This result implies that the crests [troughs] in the mismatch $\epsilon$ will occur at time delays $\Delta t_{d,c}$ [$\Delta t_{d,t}$] for which $w_{\rm cut} = 2n\pi$ [$(2n+1)\pi$], i.e.
\begin{subequations} \label{appendix eq:predicted troughs crests}
\begin{align} 
    \Delta t_{d,c} &= nf_{\rm cut}^{-1} = n \times 10.47\,{\rm ms} \left( \frac{M_z}{45.9 M_\odot} \right), \label{E:CP} \\
    \Delta t_{d,t} &= \left( n + \frac{1}{2} \right)f_{\rm cut}^{-1} = \left( n + \frac{1}{2} \right)10.47\,{\rm ms} \left( \frac{M_z}{45.9 M_\odot} \right). \label{E:TP}
\end{align}
\end{subequations}

\begin{table}[t!]
    \centering
    \begin{tabular}{ |P{1.0cm}|P{1.5cm}|P{1.5cm}|P{1.5cm}|P{1.5cm}| }
    \hline
    & \multicolumn{2}{|c|}{Crests} & 
    \multicolumn{2}{|c|}{Troughs} \\
    \hline
    $n$ & Numerical $\Delta t_{d}$ [ms] & Predicted $\Delta t_{d,c}$ [ms] & Numerical $\Delta t_{d}$ [ms] & Predicted $\Delta t_{d,t}$ [ms]\\
    \hline
    1  & 11.51 &  10.47 & 15.91 &  15.71\\  
    2  & 21.31 &  20.94 & 26.12 &  26.18\\ 
    3  & 32.52 &  31.41 & 36.72 &  36.65\\
    4  & 42.13 &  41.88 & 47.33 &  47.12\\
    5  & 53.53 &  52.36 & 57.13 &  57.59\\
    6  & 63.54 &  62.83 & 68.34 &  68.06\\
    7  & 73.75 &  73.30 & 78.15 &  78.53\\
    8  & 84.75 &  83.77 & 88.95 &  89.01\\
    9  & 94.36 &  94.24 & 99.56 &  99.48\\
    10 & 105.8 &  104.7 & 109.6 &  109.9\\
    11 & 115.4 &  115.2 & 120.8 &  120.4\\
    12 & 126.2 &  125.7 & 130.6 &  130.9\\
    13 & 136.6 &  136.1 & 141.8 &  141.4\\
    14 & 146.6 &  146.6 & 151.8 &  151.8\\
    15 & 157.8 &  157.1 &   -   &     - \\
    \hline
    \end{tabular}
    \caption{
    The five columns in this table list: (1) the index $n$ of the crests and troughs, (2) the numerically determined time delay of the crest, (3) the time delay of the crest predicted by Eq.~(\ref{E:CP}), (4) the numerically determined time delay of the trough, and (5) the time delay of the trough predicted by Eq.~(\ref{E:TP}).  The lens and source parameters are the same as those in Fig.~\ref{fig:mismatch_mcz_20_I_0.2}.
    }
    \label{tab:crest_troughs_mcz_20}
\end{table}

Fig.~\ref{fig:mismatch_mcz_20_I_0.2} reproduces the blue curve shown in the bottom left panel of Fig.~\ref{fig:1d mismatch y-ML} corresponding to the mismatch $\epsilon$ between a lensed GW source and unlensed templates in the SIS model for flux ratio $I = 0.2$ and source parameters $\mathcal{M} = 20 M_\odot$, $\eta = 0.25$, and $t_c = \phi_c = 0$. The dashed (dotted) vertical lines indicate the values of the time delay at which maxima (minima) of the mismatch occur. These numerically determined values are listed in Table~\ref{tab:crest_troughs_mcz_20}, where they are compared to the values predicted by Eqs.~(\ref{E:CP}) and (\ref{E:TP}).  We see that there is excellent agreement between our numerical results and analytical predictions, and that this agreement improves for larger times delays $\Delta t_d$ at which the extreme geometrical-optics approximation becomes increasingly valid.

\bibliography{bibme}

\end{document}